\documentclass[twocolumn,showpacs,preprintnumbers,amsmath,amssymb,prb]{revtex4}

\usepackage{graphicx}
\usepackage{dcolumn}
\usepackage{bm}
\begin{document}
\title{Charge disproportionation in (TMTTF)$_{2}$SCN observed by $^{13}$C NMR}
\author{Shigeki Fujiyama}
\email{fujiyama@ims.ac.jp} 
\author{Toshikazu Nakamura}
\affiliation{Institute for Molecular Science, Okazaki 444-8585, Japan}
\date{\today}

\begin{abstract}
The results of $^{13}$C NMR spectra and nuclear spin-lattice relaxation rate $1/T_1$ for the quasi-one-dimensional quarter-filled organic material (TMTTF)$_2$SCN are presented.
A new inhomogeneous line appears below the anion ordering temperature ($T_{\textrm{AO}}$), below which the intensity of the distinct line owing to the inner carbon site from the inversion center is halved. This indicates that the charge transfer (disproportionation) occurs from one inner carbon site to the other inner site exclusively within one unit cell triggered by anion ordering.
The $1/T_1$ for the charge-donating inner site shows activated temperature dependence below $T_{\textrm{AO}}$, although $1/T_1$ for the outer carbon decreases much more moderately in lowering temperature as has been discussed in weakly interacting one dimensional conductors.
Our observation of the local gap for the spin excitation generated by anion ordering is consistent with the simple model assuming only the electrostatic interaction between the inner carbon sites and ordered anions, which is known to be much closer to the inner sites than to the outer sites.
Nevertheless, we reveal that only the electrostatic interaction between the anions and molecules is insufficient to reproduce the observed divergence of the frequency shift and the linewidth of the inhomogeneous line stemming from the charge-accepting inner site at a much lower temperature than $T_{\textrm{AO}}$.
\end{abstract}

\maketitle
\section{Introduction}
\label{sec:Intro}
There has been considerable interest in quasi-one-dimensional (Q1D) correlated electrons. Theoretical studies on generalized Hubbard models for Q1D electronic systems have revealed a rich phase diagram including various instabilities towards spin-Peierls, antiferromagnets, spin density wave (SDW) and superconductivity.~\cite{Schultz1994,Kishine2002}

The physical properties of molecular based Q1D quarter-filled materials, (TMT$C$F)$_{2}$\textit{X} ($C$=Se (S)), also known as the Bechgaard salts (and their sulphur analogues), have been extensively studied so far because the materials realize various ground states by modifying calcogens ($C$) and anions ($X$).~\cite{Ishiguro1998} The macroscopic electronic properties of (TMT$C$F)$_{2}$\textit{X} are summarized in a pressure vs. temperature phase diagram.~\cite{Jerome1991}

The resistivities of the (TMTTF)$_2$\textit{X} ($C$=S)  are two orders higher than those of (TMTSF)$_2$\textit{X} ($C$=Se). The temperatures where the resistivities show their minima are about 200 K, which is one order higher than those for TMTSF materials.~\cite{Delhaes1979,Laversanne1984} In addition, the ground states of the TMTTF family are driven by magnetic instabilities undergoing spin-Peierls or antiferromagnetic phase transitions, in spite that those of TMTSF are driven by Fermi surface instabilities towards incommensurate SDW or superconductivity. This implies that the strong electronic correlation becomes relevant for the electronic states of the TMTTF materials at low temperatures.

Recently, it is argued that the electronic charges are disproportionated along the chain direction by measuring dielectric response,~\cite{Nad1998} XRD (X-ray diffraction),~\cite{Pouget1997} and $^{13}$C NMR spectra~\cite{Chow2000} in the semiconducting temperature region. Theoretical studies on quarter-filled-Q1D electronic systems based on the extended Hubbard models assuming onsite and intersite repulsions predict several electronic instabilities toward $2k_F$-SDW, $2k_F$-charge density wave (CDW), and $4k_F$-CDW including their coexistences.~\cite{Seo1997,Kobayashi1997,Ung1994,Mazumdar1999}

The ground state of (TMTTF)$_{2}$SCN is revealed to be the commensurate antiferromagnet by Coulon \textit{et al.}~\cite{Coulon1982} The dependence of the $^{1}$H-NMR spectra on the direction of the external magnetic field suggests that nodes of spin density exist on the TMTTF molecules.~\cite{Nakamura1997,Nakamura1995} Seo and Fukuyama proposed a periodic (up-0-down-0) magnetic structure for the antiferromagnetic state in this material.~\cite{Seo1997} They also pointed out that the intersite repulsive interaction is indispensable to reproduce the charge disproportionated ground state in mean field calculations, which concludes $4k_F$-CDW state at low temperatures.

In this report, we demonstrate the charge disproportionation in (TMTTF)$_2$SCN by the $^{13}$C NMR spectra, which is triggered by anion ordering. The contrasting temperature dependence of the linewidth of the spectrum and nuclear spin lattice relaxation rates $1/T_1$ for distinct lines indicates that the anion ordering generates local and random potentials at the charge-donating inner carbon site. Our observations also suggest that charge transfer occurs within one unit cell in contrast to (TMTTF)$_{2}$\textit{M}F$_6$ (\textit{M}=As, P) that have spin-Peierls ground states. We argue the charge configuration below $T_{\textrm{AO}}$ by comparing the results of XRD. The model assuming that only source for the charge disproportionation is the electrostatic interaction between the ordered anions and inner carbon sites that is much closer to the anions than the outer sites can leads to the $4k_F$-CDW charge configuration which includes the local and random electronic state.

However, the divergence of the frequency shift and the linewidth of the inhomogeneous line stemming from the charge-accepting inner site at a temperature much lower than $T_{\textrm{AO}}$ needs other source to generate the slowing down of charge fluctuations than the electrostatic interaction between the anions and molecules. 

We propose the instability toward the CDW which has a commensurate wavelength originated from the electronic correlation among the molecules as a candidate that can make a frustrated electronic state between the disordered state owing to the imperfect anion ordering.

\section{Experimentals}
\label{sec:Exp}
\begin{figure}[htp]
\centering
\includegraphics*[width=6cm]{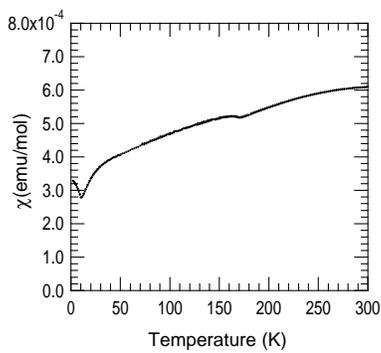}
\caption{Temperature dependence of the uniform susceptibility measured by SQUID at the external field of 5.0 T.}
\label{fig:ESR}
\end{figure}

A rectangular plate-like single crystal of (TMTTF)$_{2}$SCN in which the two central carbon sites on the TMTTF molecules are labeled with $^{13}$C was prepared by the standard electrochemical oxidation method. The uniform susceptibility of our sample is shown in Fig.\ \ref{fig:ESR}.
As shown in Fig.~\ref{fig:ESR}, the uniform susceptibility ($\chi$) in the paramagnetic state gradually decreases in lowering temperature with a small dip at $T_{\textrm{AO}}\sim 170$ K. Below 40 K, the $\chi$ shows stronger decrease than that above 40 K.

The NMR measurements were performed by using a standard pulsed NMR spectrometer operated at 87.12 MHz for the single crystal. The spectra were obtained by the Fourier transformation of the solid echo refocused by applying a pair of $\pi/2$ pulses shifted in phase by $\pi/2$. The nuclear recovery data were measured by the saturation recovery method, and $1/T_1$ were obtained by fitting them to a single exponential formula.

\section{$^{13}$C NMR spectra}
\label{sec:Shift}
\begin{figure}[htp]
\centering
\includegraphics*[width=7cm]{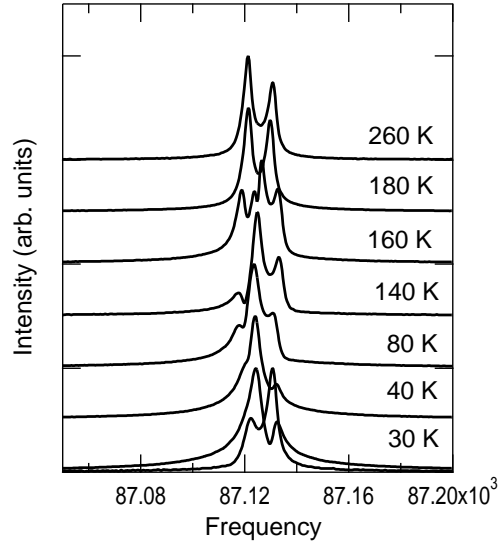}
\caption{$^{13}$C NMR spectra of (TMTTF)$_2$SCN in the paramagnetic state.}
\label{fig:spec160K}
\end{figure}

\begin{figure}[htp]
\centering
\includegraphics*[width=8cm]{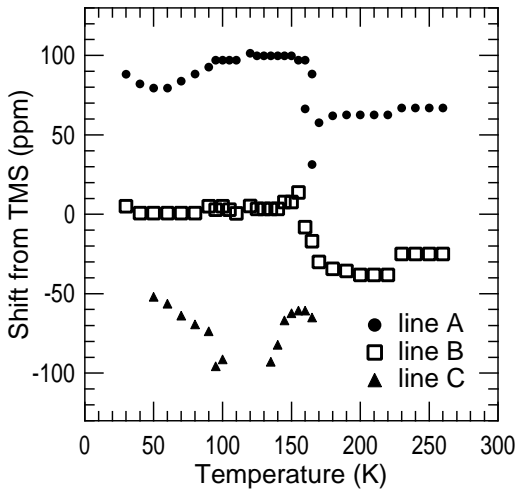}
\caption{Frequency shifts from TMS in the paramagnetic state.}
\label{fig:Shift}
\end{figure}

We show in Fig.~\ref{fig:spec160K} the $^{13}$C NMR spectra of (TMTTF)$_{2}$SCN in the paramagnetic state. We applied the external magnetic field along the \textit{magic angle} to overlap the doubly split absorption lines due to the nuclear dipolar interaction. The observed two distinct lines at room temperature correspond to two inequivalent carbon sites: one is the outer carbon site from the inversion center in the unit cell; and the other is the inner site. The molecular orbital calculation predicts that the spin density on the inner carbon site is almost 1.4 times as large as that on the outer site in the TMTTF molecules.~\cite{Ishiguro1998,Kinoshita1984} We also plot in Fig.~\ref{fig:Shift} the frequency shifts from TMS (tetramethylsilan) of the distinct lines of the spectra as a function of temperature.

According to the molecular orbital calculation, the dominating character of the central carbon site participating in the highest occupied molecular orbital is the $2p_z$ orbital, where $z$ is the direction perpendicular to the TMTTF plane. If the dipolar field from other orbitals than $^{13}$C $2p_z$ is neglected, the frequency shift $K$ is expected to have the uniaxial symmetry, $K(\theta)=K_{\textrm{iso}}+(3\cos^2\theta-1)K_{\textrm{ax}}$. Here $\theta$ is the angle between the external field and the $z$-axis, $K_{\textrm{iso}}(K_{\textrm{ax}})$ is the isotropic (anisotropic) part of the magnetic frequency shift, which is related to the spin susceptibility ($\chi$) by $K_{\textrm{iso,ax}}=F_{\textrm{iso,ax}}\chi/2N_A \mu_B$. Here, $N_A$ is the Avogadro's number, and $\mu_B$ is the Bohr magneton. The quantity $F_{\textrm{ax}}$ is proportional to the spin density in the $2p_z$ orbital as $F_{\textrm{ax}}=\frac{2}{5}<r^{-3}>_{2p}\mu_B \sigma$.~\cite{Abragam1951,Owen1966} Here $<r^{-3}>_{2p}$ is the expectation value of $r^{-3}$ for the $2p_z$ orbital of the central carbon site, and $\sigma$ is the fractional spin density in the orbital. In our experimental condition, the $z$-axis is almost perpendicular to the external field, therefore the spin density $\sigma$ in the $2p_z$ orbital of the $^{13}$C is expected to reduce the total hyperfine coupling constant, that results in the small $T$ dependence of $K$ above 180 K as shown in Fig.~\ref{fig:Shift}.

In lowering temperature, two lines split into four lines between 165 K and 160 K. However, only three lines are observed below 160 K as shown in Figs.~\ref{fig:spec160K} and \ref{fig:Shift}.

It is well known that non-centrosymmetric anions in (TMT$C$F)$_{2}$\textit{X} can take two orientations in the crystal. At room temperature, the directions of the anions are disordered, which results in the fact that the inversion centers are present on average only. Below the anion ordering temperature ($T_{\textrm{AO}}$), a superstructure is observed by XRD measurements. Pouget \textit{et al.} found that the wave vector of the anion ordering in (TMTTF)$_{2}$SCN is $\textbf{\textit{q}}=(0, 1/2, 1/2)$.~\cite{Pouget1996} This exhibits that the superstructure is not in the chain ($a$-) direction.

\begin{figure}[htp]
\centering
\includegraphics*[width=6cm]{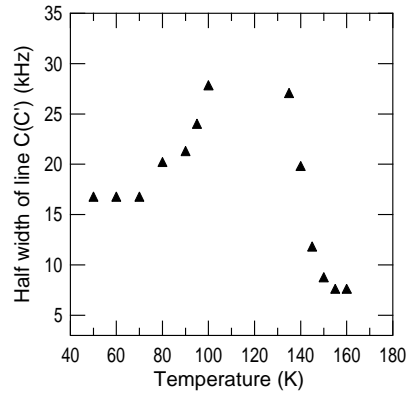}
\caption{Half width at the 0.2 point of the maximum for the line $C$($C'$).}
\label{fig:HW8M}
\end{figure}

The variation of the NMR spectra can be easily ascribed to the structural phase transition at $T_{\textrm{AO}}$. Four lines are observed in the spectra in the narrow temperature range of a transitional state between the anion disordered and ordered states. Between 135 K and 155 K, the NMR spectra are decomposed into two lines of homogeneous linewidth (lines $A$ and $B$ henceforce) and one inhomogeneously broadened line (line $C$). It should be noted that the intensity of line $A$, which is located at the highest frequency in the spectrum, is almost halved below $T_{\textrm{AO}}$. The frequency shifts of the lines $A$ and $B$ show little $T$ dependence between 50 K and 155 K, which indicates that the spin densities at the carbon sites corresponding to both lines are unchanged or less than those above $T_{\textrm{AO}}$. In contrast, the line $C$ progressively shifts to the smaller-frequency side with  inhomogeneous broadening in lowering temperature. At 130 K, the line $C$ disappears and only the lines $A$ and $B$ remain in the spectra. However, a new inhomogeneous line ($C'$) becomes visible below 100 K. As shown in Fig.~\ref{fig:Shift}, the line $C'$ shifts in the opposite manner to the line $C$ with the turning point around 120 K, and finally merges into the line $B$ at 50 K. We also plot the half width at the  20 \% maximum of the line $C$ ($C'$) in Fig.~\ref{fig:HW8M}. As shown in Fig.~\ref{fig:HW8M}, the linewidth becomes broader in decreasing temperature between 160 K and 135 K, which indicates wide distribution of the charge density for line $C$. Approaching 135 K, the width of the line $C$ diverges and finally the line $C$ disappears. Below 100 K where the line $C'$ appears, the linewidth becomes narrower in lowering temperature. 
Since the uniform susceptibility is almost constant between 50 and 170 K, the origin of the large shift and the divergence of the width of the line $C$ has to be attributed to the local variation of the electronic state at the carbon site.

\section{Nuclear spin-lattice relaxation rate}
\label{sec:T1}
\begin{figure}[htp]
\centering
\includegraphics*[width=8cm]{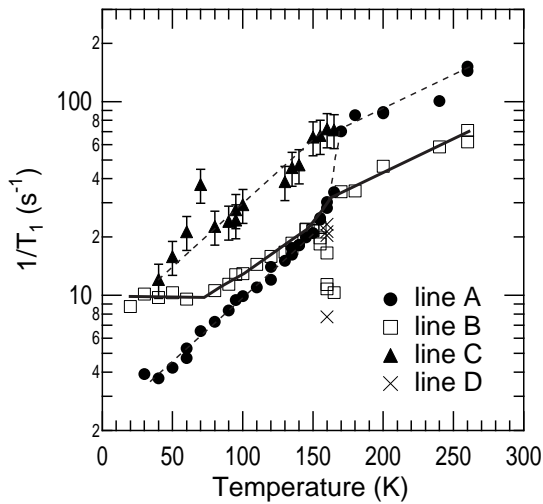}
\caption{Nuclear spin-lattice relaxation rates for distinct lines. The solid and dashed lines are guides to the eye.}
\label{fig:T1}
\end{figure}

We show in Fig.~\ref{fig:T1} the nuclear spin-lattice relaxation rate of $^{13}$C, $1/T_1$ for the distinct lines. $1/T_1$ for the line $A$ ($1/T_1^A$) and that  for the line $B$ ($1/T_1^B$) increase with temperature in a parallel manner above $T_{\textrm{AO}}$ in the logarithmic scale as shown in Fig.~\ref{fig:T1}. $1/T_1$ at site $n$ is expressed as, $1/T_1^n=\frac{2\gamma_n^2T}{\mu_B^2}\sum_q \: ^n F_\perp(q)^2 \chi''(q,\omega_L)/\omega_L$, here $\gamma_n$ is the gyromagnetic ratio of carbon, $^n F_\perp(q)$ is the $q$ dependent hyperfine coupling constant at site $n$, $\chi''$ is the imaginary                                                                                                                                                                                                                                                                                                                                                                                                                                                                                                                                                                                                                                   part of the dynamic susceptibility, and $\omega_L$ is the Larmor frequency. Thus, the only difference between $1/T_1^A$ and $1/T_1^B$ comes from that of $^n F_\perp(q)$. Since the local spin correlation function is expected to have little $q$ dependence in organic conductors, which results in uniform hyperfine coupling constants, the $F_\perp(q)$ exclusively depends on the spin density at the nucleus. Since $1/T_1^A$ appears twice as large as $1/T_1^B$ above $T_{\textrm{AO}}$, we conclude that the line $A$ in the frequency domain is due to the inner carbon site in the TMTTF molecule.

In the narrow temperature range between 160 and 170 K where four distinct lines are observed in the NMR spectra, $1/T_1^B$ (solid square) shows marked decrease. The $1/T_1^B$ at 165 K becomes almost half of $1/T_1^B$ at 170 K, but is recovered at 160 K. The small dip within a 10 K interval is also observed in the uniform susceptibility as shown in Fig.~\ref{fig:ESR}. Below 160 K, $1/T_1^B$ again decreases in lowering temperature down to 60 K and has a crossover to the low-$T$ regime where $1/T_1^B$ becomes independent of $T$. This crossover is discussed by Bourboneis and his collaborators in the framework of weak interacting Q1D electronic systems in Refs.~\onlinecite{Bourbonnais1993} and \onlinecite{Wzietek1993}.

On the other hand, $1/T_1^A$ (solid circle) changes differently from $1/T_1^B$ for $T$ below $T_{\textrm{AO}}$. $1/T_1^A$ strongly decreases from 170 K to 140 K compared with that in the metallic state. Since $1/T_1^B$ shows much more moderate $T$ dependence in this temperature range, the only source of the marked decrease of $1/T_1^A$ is the loss of hyperfine coupling $|^A F_\perp(q)|$, and the corresponding loss of charge density at site $A$.

\begin{figure}[htp]
\centering
\includegraphics*[width=6cm]{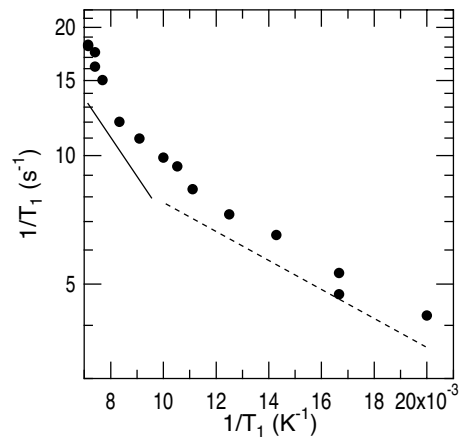}
\caption{Arrenius plot of the nuclear spin-lattice relaxation rate for the line $A$ below 150 K. The solid (dashed) line shows the activated temperature dependence with $\Delta$=200 K (70 K) by a crude fitting of the data between 140 K and 100 K (between 100 K and 40 K) to the formula, $1/T_1\propto \exp(-\Delta/T)$.}
\label{fig:iT1activ}
\end{figure}

The $1/T_1^A$ continues to decrease more steeply than $1/T_1^B$ below 140 K in lowering temperature. The ratio $T_1^A/T_1^B$ is about 0.5 above $T_{\textrm{AO}}$, but the ratio becomes 2 at 40 K. Therefore, the nuclear relaxation process below $T_{\textrm{AO}}$ for the line $A$ is concluded to be insensitive to the uniform susceptibility ($\chi$), and seems to have an activation gap for the corresponding spin excitations. We plot $1/T_1^A$ as a function of $1/T$ in the logarithmic scale in Fig.~\ref{fig:iT1activ}. The observation of the convex curve in Fig.~\ref{fig:iT1activ} indicates that the activation energy $\Delta$ decreases in lowering temperature. The solid (dashed) straight line corresponds to $\Delta$ = 200 K (70 K) in the activated $T$-dependence of $1/T_1\propto \exp(-\Delta/T)$. It should be mentioned that $\Delta$ is comparable to the temperature, indicating that the nuclear relaxation process at site $A$ is originated from the electronic correlation of thermally excited two magnons beyond the random barriers.

We also plot $1/T_1$ for the lines $C$ and $C'$ in Fig.~\ref{fig:T1} by solid triangles. The $1/T_1^C$ is three times as large as $1/T_1^A$ at 150 K, and decreases in lowering temperature in almost parallel manner with $1/T_1^A$.
It is noticeable that $1/T_1^{C'}$ follows the extrapolated line of $1/T_1^C$ between 150 K and 130 K, which indicates similar spin densities for both lines. Therefore we consider that the line $C'$ is assigned to the same carbon site with the line $C$. 

\section{Discussion}
\label{sec:discussion}
The prominent features of the frequency shifts and $1/T_1$ for distinct lines of the $^{13}$C NMR as shown in Secs.~\ref{sec:Shift} and \ref{sec:T1} can be summarized as follows. An inhomogeneous line appears below $T_{\textrm{AO}}$, which can be ascribed to the charge disproportionation at the inner carbon sites because the intensity of one inner carbon is halved. 
Second is the unexpected local fluctuation of electronic charges and the random potential at the inner carbon sites, which may not be participate in the highest occupied molecular orbital.

In this section, we discuss our observations by comparing the results of XRD~\cite{Pouget1996} to discriminate the electronic correlation in this quarter-filled Q1D system from trivial electrostatic interactions stemming from the anion ordering.

\begin{figure}[htp]
\centering
\includegraphics*[width=5cm]{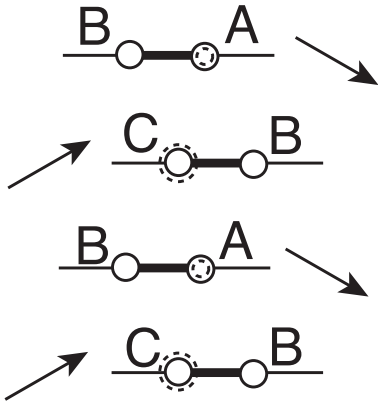}
\caption{Schematic representation of the anion configuration that realizes $\textit{\textbf{q}}$=(0, 1/2, 1/2).~\cite{Pouget1996} The arrows symbolize the short contact interaction to the TMTTF molecules. The circles represent two carbon sites in the TMTTF molecules, and the symbols $A$, $B$ and $C$ correspond to the distinct lines in the NMR spectra discussed in the text.}
\label{fig:AOXray}
\end{figure}

We now come back to the NMR spectra below $T_{\textrm{AO}}$, which have three distinct lines. Pouget \textit{et al.} found the loss of the inversion symmetry in the unit cell below $T_{\textrm{AO}}$, and proposed the configuration of SCN anions  schematically represented by the arrows in Fig.~\ref{fig:AOXray}. They argued that the anions alternately establish long and short contacts along the $a$- (chain) direction. Therefore, much more effect on the charge density at the inner carbon site by the variation of the distance between the molecule and anion is expected than that at the outer site. Indeed, our result of $1/T_1^B$ that follows the uniform susceptibility shows the smaller effect on the charge density at the outer carbon site at $T_{\textrm{AO}}$. On the other hand, the remarkable decrease of $1/T_1^A$ between 160 K and 145 K, and being three times as large as $1/T_1^A$ of $1/T_1^C$ shows that the charge disproportionation takes place exclusively at the inner carbon sites. Therefore, the charge transfer can be considered to take place from the one inner carbon site ($A$) to the other one (site $C$) triggered by anion ordering. This argument leads to the possible \textit{intra-molecular} charge disproportionation in contrast to (TMTTF)$_2$$M$F$_6$ which has spin-Peierls ground states.

If we assume that the dominating source to modify the charge densities at the inner carbon sites is the electrostatic interaction between the molecules and anions, it is expected that the carbon site which is closer to the anion ($C$ in Fig. \ref{fig:AOXray}) attains more positive charge density.

As mentioned in Sec.~\ref{sec:T1}, the lines $C$ and $C'$ are assigned to the same carbon site because of the same                                                                                                                                          magnitudes of the spin densities for both lines. We discuss here the disappearance of the line $C$ ($C'$) between 130 K and 100 K. 

The loss of the NMR intensity observed for the line $C$ is known as a "wipeout" phenomenon when the electronic state is inhomogeneous causing broadening of the NMR spectra and shortening of the spin echo decay time.~\cite{Abragam,Winter} In the low dimensional correlated electronic materials, Hunt \textit{et al.} first demonstrated the intensity loss of the Cu NQR spectrum in the two dimensional CuO$_2$ plane of La$_{2-x}$Ba(Sr)$_{x}$CuO$_{4}$, in which microscopic charge ordering is argued.~\cite{Hunt1999} For the Q1D cuprates, the intensity of the Cu NMR spectra of the hole-doped Cu$_2$O$_3$ two-leg ladder, Sr$_{14-x}$Ca$_{x}$Cu$_{24}$O$_{41}$, decreases in the semiconducting temperature regime in lowering temperature.~\cite{Fujiyama2002} In these cases, the shortening of the spin echo decay time ($T_2$) due to the strong spin fluctuation in the vicinity of charge stripes or the slow fluctuation of the electric field gradients generates the wipeout of the NMR spectra. In the present study, the most plausible origin of the wipeout is also the shortening of the spin echo decay time due to the slowing down of charge fluctuations because the frequency shift and the linewidth of the NMR spectra diverges at around 120 K. Since it is known that $1/T_2$ takes a maximum value when the inverse of the correlation time of electronic fluctuations is close to the second moment of the NMR spectrum, the fluctuation frequency of the local field is confirmed to be less than 50 kHz below 120 K.~\cite{Takigawa1986}

It is noteworthy that the critical phenomenon observed for the site $C$ occurs at a much lower temperature than $T_{\textrm{AO}}$. In addition, no significant anomaly in the macroscopic measurements such as resistivity and uniform susceptibility is reported around 120 K so far, and $1/T_1^B$ that follows the uniform susceptibility remains intact around this temperature. Therefore the significant slowing down of the electronic charges may be due to the local electronic interactions.

It is reported that the superspot at $\textit{\textbf{q}}=(0, 1/2, 1/2)$ in the reciprocal space appears at 160 K, and the intensity at this point continues to increase almost linearly down to 50 K according to XRD.~\cite{Pouget2000} Between 160 K and 50 K where wide distribution remains in the displacements of anions from the original position or the partial ordering with a finite correlation length, the electrostatic interaction between the anions and TMTTF molecules would lead to disordered charge densities at site $C$. Indeed, the observation of the activated $T$ dependence with multi-barriers in $1/T_1^A$ reveals random potentials at site $A$. Since the energy scale of XRD is much higher than the Larmor frequency ($\sim$ 100 MHz), our observation of the divergence of the linewidth observed at an intermediate temperature between the onset and saturating temperature for the anion ordering shows that it can not be concluded that the only source of the strong slowing down of the local charge fluctuation at site $C$ is the electrostatic coupling between the anions and molecules. Instead, it is plausible that some source that can compete with the disordered electronic state generates a frustration at the carbon site below $T_{\textrm{AO}}$.

It is recently argued that some members of the (TMTTF)$_2$\textit{X} family have a charge disproportionated phase of $4k_F$-CDW in the semiconducting temperature regime. Chow \textit{et al.} found the splitting of $^{13}$C NMR spectra for (TMTTF)$_2$\textit{M}F$_{6}$ ($M$=As, P), which have the spin-Peierls ground state.~\cite{Chow2000} This charge ordering phenomenon is regarded as a result of  the strong Coulomb repulsive interaction between the electrons that has spin 1/2 through the dimerization of TMTTF molecules. The observation of the spin-Peierls ground state in (TMTTF)$_{2}$\textit{M}F$_{6}$ evidences the strong dimerization in these materials. However, the XRD experiments for the SCN salts do not show an obvious superstructure along the chain direction, resulting in weak dimerization at most.~\cite{Pouget1996,NogamiUnp} Our observation of the charge disproportionation with $4k_F$-CDW like charge configuration is the first one in this weakly dimerized, quarter-filled Q1D TMTTF material.

The CDW instability which has the commensurate wavelength originated from the electronic correlation as discussed for the electronic state of (TMTTF)$_{2}$\textit{M}F$_{6}$ may be a candidate that competes with the disordered local electronic state at the inner carbon sites in the intermediate temperature range of imperfect anion ordering. We should mention that there still remains an open question on the mechanism of the competition between the local activation gap generated by the anion ordering and the instability toward the CDW originated from the electronic correlation among the molecules. Theoretical studies that include the interactions between the anions and molecules are expected.

\section{Conclusion}
We measured the $^{13}$C NMR spectra and nuclear spin lattice relaxation rate $1/T_1$ for (TMTTF)$_2$SCN that has a linear anion. The appearance of the new inhomogeneous line (line $C$) below the anion ordering temperature ($T_{\textrm{AO}}$) reveals that the electronic charges are disproportionated in this material. 

In contrast to (TMTTF)$_2$\textit{M}F$_6$ that have spin-Peierls ground states, the charge transfer in this material leads to the local fluctuations of electronic charges which induces the different $T$ dependences in the frequency shifts and $1/T_1$. Of these, the strong decrease of $1/T_1$ for the charge donating site (site $A$) in lowering temperature revealed an activation gap for spin excitations and the inhomogeneous electronic state possibly due to the wide distribution of the displacements from the original position of the anions as observed by XRD.

These results are consistent with the wavevector of anion ordering which naturally leads to $4k_F$-CDW like charge disproportionation by assuming the electrostatic interaction between the anions and molecules.

However, our observation of the divergence of the frequency shift and the linewidth of the spectra of the charge-accepting site $C$ indicating the gradual slowing down of the local electronic charges at a much lower temperature than $T_{\textrm{AO}}$ needs other source than the trivial electrostatic interaction between the anions and molecules.

We propose the instability toward the $4k_F$-CDW originated from the electronic correlation among the TMTTF molecules as a candidate that can make a frustrated electronic state between the disordered state.

\begin{acknowledgments}
One of the authors (S.\ F.) is grateful to Prof.\ Yonemitsu for the critical reading of the manuscript. We are also grateful to Prof.\ Yakushi, Prof.\ Brazovskii, Prof.\ Hosokoshi, Dr.\ Hiraki and Dr.\ Yamamoto for fruitful discussions. This work is supported by a Grant-in-Aid for Scientific Research on Priority Area, \textit{Development of Molecular Conductors and Magnets of New Functionality by Spin-Control} from the Ministry of Education, Culture, Sports, Science and Technology, and by a Grant-in-Aid for Scientific Research (No.\ 13640375) from Japan Society for the Promotion of Science.
\end{acknowledgments}

\end{document}